\documentclass{article}

\usepackage{epsfig}
\usepackage{amsmath,amsfonts,amssymb,graphics,graphicx,epsfig,color,times,bbm}

\newcommand{\1}{\mathbbm{1}}
\newcommand{\F}{\mathbbm{F}}

\newcommand{\tr}[1]{{\rm tr}\left[#1\right]}

\newcommand{\be}{\begin{equation}}
\newcommand{\ee}{\end{equation}}
\newcommand{\bea}{\begin{eqnarray}}
\newcommand{\eea}{\end{eqnarray}}
\newcommand{\ket}[1]{|#1\rangle}
\newcommand{\bra}[1]{\langle#1|}

\newcommand{\ave}[1]{\left\langle#1\right\rangle}
\newcommand{\norm}[1]{\lVert #1 \rVert}
\newcommand{\abs}[1]{\left| #1 \right|}
\newcommand{\ketbra}[1]{\ket{#1}\bra{#1}}

\newcommand{\eg}{\emph{e.g.}}
\newcommand{\ie}{\emph{i.e.}}

\renewcommand{\>}{\rangle}
\newcommand{\<}{\langle}
\newcommand{\C}{\mathbb{C}}

\newcommand{\proofend}{\hfill\fbox\\\medskip }

\newtheorem{theorem}{Theorem}
\newtheorem{proposition}{Proposition}

\newtheorem{lemma}{Lemma}

\newtheorem{corollary}{Corollary}

\newcommand{\proof}[1]{{\bf Proof.} #1 $\proofend$}

\begin{document}
\setlength{\textheight}{8.0truein}    %FOR 2ND PAGE ONWARDS

%\runninghead{Bell Inequalities from Multilinear Contractions}
   %         {A. Salles, D. Cavalcanti, A. Ac\'in, D. P\'erez-Garc\'ia and M. M. Wolf}

%\title{Bell Inequalities from Multilinear Contractions}

%\normalsize\textlineskip
\thispagestyle{empty}
\setcounter{page}{1}

%\copyrightheading{Vol.}{No.}{Year}{Page Nos.}
%\copyrightheading{0}{0}{2003}{000--000}

\vspace*{0.88truein}

%\alphfootnote

%\fpage{1}

\centerline{\bf
BELL INEQUALITIES FROM MULTILINEAR CONTRACTIONS
}
\vspace*{0.37truein}
%\centerline{\bf FOR QUANTUM INFORMATION AND COMPUTATION\footnote{Typeset the

\centerline{\footnotesize
ALEJO SALLES}
\vspace*{0.015truein}
\centerline{\footnotesize\it Niels Bohr Institute, Blegdamsvej 17, 2100 Copenhagen, Denmark}
%baselineskip=10pt
%centerline{\footnotesize\it City, State ZIP/Zone,
%ountry\footnote{State completely without abbreviations, the
%ffiliation and mailing address, including country. Typeset in 8
%t Times Italic.}}
\vspace*{10pt}
\centerline{\footnotesize 
DANIEL CAVALCANTI}
\vspace*{0.015truein}
\centerline{\footnotesize\it ICFO---Institut de Ci\`encies Fot\`oniques, E-08860 Castelldefels, Barcelona, Spain}
%baselineskip=10pt
%centerline{\footnotesize\it City, State ZIP/Zone, Country}
\vspace*{10pt}
\centerline{\footnotesize 
ANTONIO AC\'IN}
\vspace*{0.015truein}
\centerline{\footnotesize\it ICFO---Institut de Ci\`encies Fot\`oniques, E-08860 Castelldefels, Barcelona, Spain}
\baselineskip=10pt
\centerline{\footnotesize\it ICREA---Instituci\'o Catalana de Recerca i Estudis Avan\c cats, Lluis Companys 23, 08010, Barcelona, Spain}
\vspace*{10pt}
\centerline{\footnotesize 
DAVID P\'EREZ-GARC\'IA}
\vspace*{0.015truein}
\centerline{\footnotesize\it Departamento de An\'alisis Matemtico \& IMI, Universidad Complutense de Madrid, 28040 Madrid, Spain}
%baselineskip=10pt
%enterline{\footnotesize\it ICREA---Instituci\'o Catalana de Recerca i Estudis Avan\c cats, Lluis Companys 23, 08010, Barcelona, Spain}
\vspace*{10pt}
\centerline{\footnotesize 
MICHAEL M. WOLF}
\vspace*{0.015truein}
\centerline{\footnotesize\it Niels Bohr Institute, Blegdamsvej 17, 2100 Copenhagen, Denmark}
%\baselineskip=10pt
%centerline{\footnotesize\it City, State ZIP/Zone, Country}

\vspace*{0.225truein}
%\publisher{(received date)}{(revised date)}

\vspace*{0.21truein}

%% \abstracts{first paragraph}{second paragraph}{third paragraph}
%% If there is only one paragraph, just keep the second and third empty 
%% like the following one 
\abstract{
%%%%%%%%%%%%%%%%%%%%
% put abstract here
%%%%%%%%%%%%%%%%%%%%
We provide a framework for Bell inequalities which is based on multilinear contractions. The derivation of the inequalities allows for an intuitive geometric depiction and their violation within quantum mechanics can be seen as a direct consequence of non-vanishing commutators. The approach is motivated by generalizing recent work on non-linear inequalities which was based on the moduli of complex numbers, quaternions and octonions. We extend results on Peres' conjecture about the validity of Bell inequalities for quantum states with positive partial transposes. Moreover, we show the possibility of obtaining unbounded quantum violations albeit we also prove that quantum mechanics can only violate the derived inequalities if three or more parties are involved. 
}{}{}

\vspace*{10pt}

%\keywords{Bell Inequalities, Multilinear Contractions, Peres' Conjecture.}
\vspace*{3pt}
%\communicate{}

%\vspace*{1pt}\textlineskip    %) USE THIS MEASUREMENT WHEN THERE IS
   %) A SECTION HEADING
%\vspace*{-0.5pt}
%\noindent

%\tableofcontents

\section{Introduction}

Bell inequalities give a precise meaning to the statement that quantum correlations can be stronger than classical correlations---they represent the limits for what can be described within the framework of any local hidden variable (LHV) theory. While initially invented (by Bell, in fact~\cite{bell}) to resolve an old dispute between Einstein, Podolsky, Rosen~\cite{epr} and Bohr, they are nowadays a central tool in quantum information theory~\cite{WWReview}: Bell inequalities provide means for detecting entanglement as well as for certifying secret correlations and their violation is synonymous to usefulness in the context of quantum communication complexity~\cite{qcompl}.

There is a standard way of constructing Bell inequalities~\cite{WWReview}: first, one fixes the number of parties, observables and outcomes, all these numbers being finite; then, in the space spanned by all the observable distributions of outcomes, one considers the region accessible by LHV models; and finally one chooses a hyperplane which separates this space into two halves, one of which contains the LHV-accessible region. The linear inequality corresponding to this hyperplane is a Bell inequality and since the LHV-accessible region is a polytope, it is completely specified by a finite number of such inequalities.

As clean as this construction seems to be, it entails some drawbacks: (i) it provides no information about possible quantum violations, (ii) the inequalities are highly dependent on \eg, the number of measurement outcomes, (iii) the number of inequalities explodes with the number of parties/observables/outcomes (iv) the method does not allow to construct Bell inequalities for observables with continuous outcomes, and more vaguely (v) it is not exactly `intuitive'.

Among the various attempts to overcome some of these drawbacks our focus lies on recently derived non-linear inequalities, which are phrased in terms of moment inequalities for arbitrary observables~\cite{CFRD,SV}. An additional appealing feature of these inequalities is that their violation in quantum mechanics can be `seen' as a consequence of non-commutativity.

\begin{figure}[t!]
 \centering
 \includegraphics[width=0.5 \textwidth]{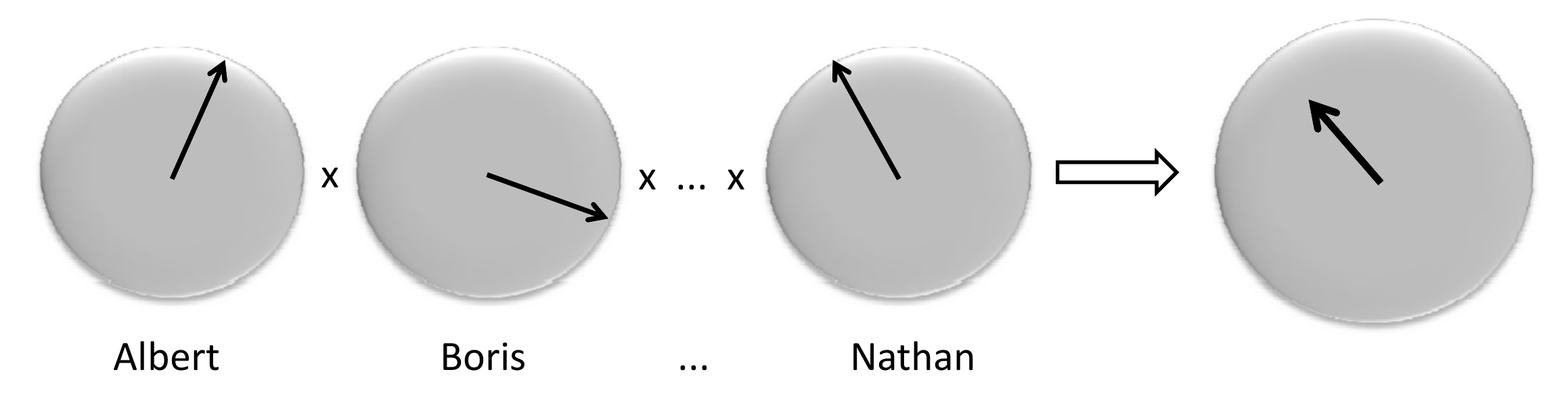}
\parbox{4in}{\caption{\footnotesize Assume Albert, Boris, ..., Nathan each have $M$ observables at hand. If we assign values to the $M$ observables on each site and arrange them into a vector in $\mathbb{R}^M$, then we can apply any multilinear contraction to $\mathbb{R}^M\times\ldots\times\mathbb{R}^M$ such that (by contractivity) the image of $N$ unit spheres on the l.h.s.\ must be contained within the unit ball on the r.h.s.\ However, if the correlations between the parties are quantum (\ie, we falsely assigned values to the observables in a LHV manner), then the 'contraction' can surprisingly expand.
\label{fig1}}}
\end{figure}

The first goal of the present work is to show that these inequalities can be imbedded in a much wider framework in which they %inequalities
can be easily understood in geometric terms---not based on the unique properties of normed division algebras which were used in the original derivations. This new framework is provided by \emph{multilinear contractions} (see Fig.\ref{fig1}).

We will give a detailed investigation of the power and limitation of this approach and thereby derive new results about the inequalities in~\cite{CFRD,SV}. Among other things we prove Peres' conjecture~\cite{Peres} %in this more general framework, i.e., that no PPT state violates any of these inequalities, and
for a subclass of the inequalities arising in this more general framework, namely, that no quantum state with a positive partial transpose (PPT) violates any inequality derived from a norm-preserving map. This proof of the conjecture applies to all the moment inequalities derived in Refs.~\cite{CFRD,SV}. We
show that unbounded quantum violations can be obtained but we also prove that generally violations only occur if three or more parties are involved. The latter can be overcome when the contractivity of the maps is stated using norms other than the Euclidian%norms different from the Euclidean norm are considered
---in fact, for the supremum norm we recover all standard correlation Bell inequalities---but then much of the new appeal of the approach is lost again. On the side we derive a variety of new results on multilinear contractions and moment inequalities which might be of independent interest.

The manuscript is organized as follows. In section~\ref{sec:Framework} we introduce the framework and derive the Bell inequalities. In section~\ref{sec:Violations} we provide examples of violations, both explicit and unbounded ones in a non-constructive fashion. We present results concerning impossibility of violation in section~\ref{sec:NoViolations}: the Peres' conjecture, the bipartite case, and that of orthogonal transformations. In section~\ref{sec:MapConstruction} we deal with the actual construction of the multilinear maps. We discuss extensions to other norms and domains in section~\ref{sec:Other}. Some useful moment inequalities are presented in section~\ref{sec:MomentInequalities}, and we finally close with a discussion in section~\ref{sec:Discussion}.

%%%%%%%%%%%%%%%%%%%
%
%
\section{Framework Construction}
\label{sec:Framework}

In order to establish our framework, we start by presenting contractive multi-linear maps, which are at the heart of the construction, and then derive the Bell inequalities from them. Subsequently, we consider some examples of inequalities arising from norm-preserving maps.

\subsection{Contractive multi-linear maps}
We introduce real multi-linear maps $T:\mathbb{R}^{M_1}\times\ldots\times\mathbb{R}^{M_N}\rightarrow \mathbb{R}^D$ (see Fig.~\ref{fig1} for a graphical depiction of the following).  For ease of notation we set all $M_k=M$ equal, but this will nowhere be required.
We say that $T$ is a \emph{contraction} if
\bea ||T(A)||&\leq& \prod_{k=1}^N ||A^{(k)}||,\label{eq:contraction}\quad \mbox{with} \ \\
\label{eq:TA}T(A)_i &=& \sum_{s} T_{i,s_1,\ldots,s_N}\prod_{k=1}^N A_{s_k}^{(k)},\eea
for all sets of vectors $\{A^{(k)}\in\mathbb{R}^{M}\}_{k=1,\ldots,N}$, and we stick for now to the Euclidean norm. Such a $T$ maps $N$ unit-spheres of $\mathbb{R}^M$ linearly into the unit-ball of $\mathbb{R}^D$. We will call it \emph{norm-preserving} if it maps these onto the unit-sphere in $\mathbb{R}^D$, \ie, if we have equality in Eq.(\ref{eq:contraction}). Norm-preserving maps are extreme points within the convex set of contractions.

\subsection{Bell inequalities based on moments}
In order to derive Bell inequalities (\ie, ones which have to be satisfied by all LHV models) from multi-linear contractions, consider a setup of $N$ parties each of which can choose out of $M$ observables. $A_j^{(k)}\in {\mathbb R}$ shall then denote the value assigned to the $j$'th observable at the $k$'th site for a single event. By taking the expectation value of the square of Eq.(\ref{eq:contraction})  and exploiting the positivity of the variance of each of the $T_i$ on the l.h.s., $\langle T_i^2 \rangle \geq \langle T_i \rangle^2$ (\ie, the convexity of the square function), we arrive at:
\begin{proposition}[Bell inequalities from contractions]\label{prop:VB} Let $T$ be a multi-linear contraction w.r.t.\ the Euclidean norm. Then  every LHV model has to satisfy
\be\label{eq:VB}  \Big|\Big|\big\langle T(A)\big\rangle\Big|\Big|^2 \leq \Big\langle \prod_{k=1}^N\big|\big| A^{(k)}\big|\big|^2\Big\rangle .\ee
\end{proposition}
Some remarks on Prop.~\ref{prop:VB} before we proceed. First note that before applying the convexity inequality, the multi-linear contraction already defines an inequality satisfied by all LHV models. However, the l.h.s.\ of Eq.~\eqref{eq:contraction} would give rise to expectations of products, which might be {\em a priori} not simultaneously measurable in a quantum setup. Second, recognize the usual LHV assumptions: (i) we assign values to the observables independent of whether they are measured (HV-assumption) and (ii) these values do not depend on the settings at the other sites (L-assumption). Third and roughly speaking the interesting maps (yielding inequalities violated by quantum mechanics) are those which are contractive but not `completely contractive', in the sense of not defining a contraction when numbers are replaced by hermitian operators. Once again, this relates the subject of Bell inequalities to the mathematical fields of tensor norms and operator spaces, which will be discussed in more detail in Sec.~\ref{sec:Unbounded}. Finally, contrary to most of the existing constructions, the inequalities are valid for an arbitrary number of outcomes, as this parameter does not play any role in the framework.

\subsection{Inequalities from norm-preserving maps}
Intuitively, the best one can do for maximizing the violation is to start from maps that are norm-preserving, so that the `gap' between the l.h.s.\ and the r.h.s.\ that has to be closed by quantum theory comes only from the convexity of the power function, and not from the contractivity of the map. Finding useful norm-preserving maps, though, proves to be a delicate endeavor.

The easiest way would be to choose $D=\prod_k M_k$ and $T$ any orthogonal transformation on $\mathbb{R}^D$, but the resulting inequalities turn out to be valid within quantum mechanics as well (see Prop.\ref{prop:useless2} below). For smaller $D$ the norm-preserving property becomes more and more restrictive. In fact, there is an entire branch of mathematics dealing in particular with the question about the smallest $D$ for which a norm-preserving $T$ exists---especially for $N=2$, \ie, quadratic forms \cite{Shapiro} which often run under the name \emph{orthogonal multiplications}. For $D=M=2,4,8$ equality in Eq.(\ref{eq:contraction}) can then easily be expressed as multiplicativity of moduli for complex numbers, quaternions and octonions. The resulting variance Bell inequalities are those presented in \cite{CFRD} ($M=2$) and \cite{SV} ($M=4,8$). More specifically, if we regard $T_{i,kl}$ as a set of $D$ real $M\times M$ matrices we can  express it in terms of Pauli matrices such that for $D=M=2$ , $T=(\1,i\sigma_y)$ is a matrix representation of complex numbers and for $D=M=4$
\be\label{eq:quat}
T=\big(\1,i \sigma_y\otimes\sigma_z,i \1\otimes\sigma_y,i\sigma_y\otimes\sigma_x\big)
\ee is a matrix representation of quaternions \cite{quaternions}.

Values of $M$ different from 2,4,8 require $D>M$ but still a large variety of orthogonal multiplications are known for relatively small $D$, \eg, from representations of Clifford algebras \cite{Clifford} or intercalate matrices \cite{intercalate}. In section~\ref{sec:NormPreserving} we provide a construction method for all norm-preserving maps.

For $M=3$, we can easily construct a norm-preserving map by noticing that the cross product $A \times B$ of two vectors in $\mathbb{R}^3$ gives another vector in $\mathbb{R}^3$ with smaller or equal norm. This is thus a contractive map, which in this case can be `dilated' to a norm-preserving one by adding to the target vector one component with the inner product $\langle A, B \rangle$ of the two original vectors. We thus get the norm-preserving {\em dilated cross product} map $T(A,B):\mathbb{R}^3 \times \mathbb{R}^3 \rightarrow \mathbb{R}^4$ given by:
\be
\label{eq:DilatedCrossProductMap}
T(A,B)=\langle A, B \rangle \oplus A \times B.
\ee
We will come back to the subject of dilations in section~\ref{sec:Contractions}.

For the multipartite case ($N>2$) the simplest construction is to concatenate orthogonal multiplications like $T_{i,klm}:=\sum_j T'_{i,jk} T''_{j,lm}$, the norm preserving property is then inherited from the building blocks. The multipartite inequalities of \cite{CFRD,SV} fall in this class. In the next section, we will make use of the inequality derived from the dilated cross product map and this concatenating procedure in order to provide an explicit example of violation.

%%%%%%%%%%%%%%%%%%%%%%%
%
%
\section{Violations}
\label{sec:Violations}

We now turn to finding violations of the moment Bell inequalities~\eqref{eq:VB}. An explicit example of violation by  10 parties was provided already in~\cite{CFRD}, whose inequality is just a specific instance of our framework for a norm-preserving map with $M=2$ and $D=2$, as we saw before. Improved examples of violations were later provided in~\cite{QCRD}. Here, we present an explicit example of violation for an entirely new inequality involving the dilated cross product map, as well as a proof that unbounded violations of the inequalities are possible. Note that for norm-preserving maps the source of possible violations by quantum mechanics becomes clear: non-commutativity. In fact, if we consider a norm-preserving map, then deriving equality in Eq.(\ref{eq:contraction}) requires cancelation and identification of terms of the form $A_i^{(k)}A_j^{(k)}$ and $A_j^{(k)}A_i^{(k)}$. If these are no longer numbers but operators, we may still follow the same algebra, however, at the cost of additional commutators popping up---terms eventually not covered by the inequality.

\subsection{Explicit violations}
\label{sec:ExplicitViolation}
The inequality we will use arises from the map constructed by concatenating two dilated cross product maps~\eqref{eq:DilatedCrossProductMap} ($\mathbb{R}^3\times\mathbb{R}^3\rightarrow\mathbb{R}^4$ each) together with a quaternion moduli product one~\eqref{eq:quat} ($\mathbb{R}^4\times\mathbb{R}^4\rightarrow\mathbb{R}^4$). By plugging the output of the two dilated cross product maps into the quaternion one, we are left with a map $T(A,B,C,D): \mathbb{R}^3\times\mathbb{R}^3\times\mathbb{R}^3\times\mathbb{R}^3\rightarrow\mathbb{R}^4$, which is norm-preserving by virtue of all the maps involved in its construction being also norm-preserving. This thus gives us an inequality for $N=4$ and $M=3$.

A state that gives a violation is the 4-party GHZ state $\ket{\psi}=\frac{1}{\sqrt{2}}(\ket{0000}+\ket{1111})$. We use spin measurements in the $xy$ plane: $\sigma(\theta)=\cos\theta \sigma_x + \sin\theta \sigma_y$. Finally, the classical bound is given by $M^N=3^4=81$.

With settings:
\begin{equation}
\begin{split}
\theta_A&=\{0,\pi/3,2\pi/3\}\\
\theta_B&=\{0,-\pi/3,-2\pi/3\}\\
\theta_C&=\{0,\pi/3,-\pi/3\}\\
\theta_D&=\{0,-\pi/3,\pi/3\},\\
\end{split}
\end{equation}
where different letters label the parties, and for each we give the three angles they choose among, we get a violation $324>81$.

\subsection{Unbounded violations}
\label{sec:Unbounded}

The mathematical theory of Operator Spaces turned out to be a useful tool to deal with standard Bell inequalities \cite{operatorspaces, unbounded} when it comes to questions about the asymptotic scaling of the Bell inequality violation. Here we show how it also helps in the present situation to produce unbounded violations in the tripartite setting---a violation which grows linearly with the dimension $n$ of the smallest Hilbert space if $n^2$ observables are used. For that we restate here a result from \cite{unbounded}:

\begin{theorem}
For every $n\in \mathbb{N}$ there exist $d\in \mathbb{N}$, a complex trilinear contraction $T:\C^{n^2}\times \C^{d^2}\times \C^{d^2}\longrightarrow \C$ and a $n\times d\times d$ pure state $|\psi\>$ such that
$$\frac{1}{\sqrt{n}d}\left|\sum_{rr'ss'tt'} T_{rr'ss'tt'}\<\psi|rst\>\<r's't'|\psi\>\right|\ge C\sqrt{n}\; .$$
$C$ is a universal constant and $|\psi\>$ has an explicit form in terms of $d\times d$ random unitaries: $$|\psi\>=\frac{1}{\sqrt{nd}}\sum_{ijk} \<j|U_i|k\>\; |ijk\>\; .$$
\end{theorem}

The pure state and the contraction appearing in the previous theorem characterize the quantum state and Bell inequality leading to an unbounded violation. Now, we can actually specify the local measurements by the parties. In order to do that, we decompose any matrix element $|a\>\<a'|$, where $a$ and $a'$ label the elements of the computational basis, in real and imaginary part as $$|a\>\<a'|=\frac{|a\>\<a'|+|a'\>\<a|}{2}+i\frac{i|a'\>\<a|-i|a\>\<a'|}{2}\; . $$ This gives us sequences of observables $A_{rr'}, B_{ss'}, C_{tt'}$  where either $A_{rr'}=\frac{|r\>\<r'|+|r'\>\<r|}{2}$ for all $r$ or  $A_{rr'}=\frac{i|r'\>\<r|-i|r\>\<r'|}{2}$ for all $r$ (and similarly for $B_{ss'},C_{tt'}$), for which
\bea && \<\psi|\sum_{rr'ss'tt'} T'_{rr'ss'tt'}A_{rr'}\otimes B_{ss'}\otimes C_{tt'} |\psi\>^2\ge \nonumber \\ &&
\ge\frac{C^2n}{2^8}\|\sum_{rr'}A_{rr'}^2\|\|\sum_{ss'}B_{rr'}^2\|\|\sum_{tt'}C_{tt'}^2\|\ge\nonumber \\ &&
 \ge \frac{C^2n}{2^8} \<\psi|\sum_{rr'ss'tt'}A_{rr'}^2\otimes B_{ss'}^2\otimes C_{tt'}^2 |\psi\>\; .\label{eq:unboundedn}\eea
Here $T'$ is either the real or the imaginary part of $T$. Both are clearly {\it real} trilinear contractions.
Note that for the first inequality  we were using $$\|\sum_{rr'}A_{rr'}^2\|\le n,\|\sum_{ss'}B_{rr'}^2\|\le d,\|\sum_{tt'}C_{tt'}^2\|\le d$$ in operator norm. Eq.(\ref{eq:unboundedn}) is the desired unbounded violation of order $n$. Note that while the observables involved are explicit (and simple) and also the states are essentially known, the present proof does not specify $T$ (i.e., the inequality) and similarly gives no control about $d$.

%%%%%%%%%%%%%%%%%%%%%%%
%
%
\section{No Violations}
\label{sec:NoViolations}

In this section, we present results concerning the impossibility of violation for certain specific instances of the inequalities. We first prove Peres' conjecture~\cite{Peres} for all inequalities derived from norm-preserving maps: these cannot be violated by any PPT state. This extends the result presented for the CFRD inequality~\cite{CFRD} for quadrature measurements in~\cite{salles}, and independently generalizes that of~\cite{nha}, where Peres' conjecture was proved in general for this specific inequality. In particular, it also proves the conjecture for the inequalities in~\cite{SV}. Second, we consider the impossibility of violation of any bipartite inequality in the framework, and lastly we study the case of maps that are orthogonal transformations, from which also no violation can arise.

\subsection{Peres' conjecture}

We will now move on to quantum systems characterized by density operators $\rho$ acting on an $N$-fold tensor product Hilbert space and observables described by Hermitian operators $A_j^{(k)}$ which act non-trivially only on the $k$'th tensor factor.  We prove the following:
\begin{proposition}[Peres' conjecture]\label{prop:Peres}
Every $N$-partite quantum state which has positive partial transpose w.r.t.\ all bipartitions satisfies all Bell inequalities arising from norm-preserving maps.
\end{proposition}

In order to show this we start with a norm-preserving map $T(A)$, with equality in Eq.~\eqref{eq:contraction}.  As noted before, when writing this down in terms of operators instead of numbers, we may always bring the l.h.s.\ into the form of a product of squares, but various steps in the algebraic transformation will have to be `paid' with a commutator. We thus have:
\be
||T(A)||^2 = \prod_{k=1}^N\big|\big| A^{(k)}\big|\big|^2+ \mathcal O,
\label{eq:commutatorscomplete}
\ee
where $\mathcal O$ encompasses the commutator terms, and is given by:
\be
\mathcal O = \sum_t \big[A_{\sigma}^{(\kappa)},A_{\sigma'}^{(\kappa)}\big]\otimes R_t.
\label{eq:odef}
\ee
Here, the index $t$ labels the steps needed in order to derive the norm-preserving property. The operator $R_t$ just denotes the remainder at each step of the computation and the site $\kappa$ as well as the observable labels $\sigma,\sigma'$ of course depend on $t$. Without specifying $T$ (not to speak about the algebraic transformation) there is not much to say about the $R_t$'s. The coarse structure in Eqs.~\eqref{eq:commutatorscomplete} and~\eqref{eq:odef}, however, suffices to prove what we want. At this point, the proof follows the idea in~\cite{WW00}.

Taking the expectation value in~\eqref{eq:commutatorscomplete} and applying the variance inequality, we have that, for {\em arbitrary quantum states},
\be
\Big|\Big|\big\langle T(A)\big\rangle\Big|\Big|^2 \leq \Big\langle \prod_{k=1}^N\big|\big| A^{(k)}\big|\big|^2\Big\rangle +\langle \mathcal O \rangle.
\label{eq:normpresbound}
\ee
We further note that this is true not only for arbitrary states, but also for arbitrary observables. We can thus take a state $\rho$ that is PPT according to all bipartitions $\tau$, together with the fact that $A_{s_k}^{(k) T}$ is a legitimate observable (provided $A_{s_k}^{(k)}$ is one, too) to rewrite Eq.~\eqref{eq:normpresbound} for state $\rho^{T_{\tau}}$ and observables $A^{T_\tau}$:
\be
\Big|\Big|\big\langle T(A^{T_\tau})\big\rangle_{\rho^{T_{\tau}}}\Big|\Big|^2 \leq\Big\langle \prod_{k=1}^N\big|\big| A^{(k)T_{\tau}}\big|\big|^2 \Big\rangle_{\rho^{T_{\tau}}} +\langle \mathcal O_{T_{\tau}} \rangle_{\rho^{T_{\tau}}},%\prod_{k'\notin\tau}\big|\big| A^{(k')}\big|\big|_p^p \Big\rangle_{\rho^{T_{\tau}}}
\label{eq:normpresboundTtau}
\ee
abusing the transposition notation. Here, $\tau \subseteq\{1,\ldots, N\}$ is the subset of parties that we choose to transpose, and hence labels a bipartition.

We now wish to average Eq.~\eqref{eq:normpresboundTtau} over all choices of the bipartition $\tau$, so we consider how the different terms change under the transposition operation. Stemming from $T(A)$'s linearity, the l.h.s.\ remains unchanged under transposing observables and state:
\be
\label{eq:TInvarianceLinear}
\Big|\Big|\big\langle T(A^{T_\tau})\big\rangle_{\rho^{T_{\tau}}}\Big|\Big|^2 = \Big|\Big|\big\langle T(A)\big\rangle\Big|\Big|^2.
\ee
Using that $[(A^T)^2]^T=A^2$ for any operator $A$, we see that also the first term in the r.h.s.\ is unmodified:
\be
\label{eq:TInvarianceSquares}
\Big\langle \prod_{k=1}^N\big|\big| A^{(k)T_{\tau}}\big|\big|^2 \Big\rangle_{\rho^{T_{\tau}}}=\Big\langle \prod_{k=1}^N\big|\big| A^{(k)}\big|\big|^2 \Big\rangle.
\ee
Finally, for the term involving the $\mathcal O$, we note that the terms in the sum over $t$ pick up a sign whenever $\kappa \in \tau$, given that $[A^T,B^T]^T=-[A,B]$ for operators $A$ and $B$. Similarly, $R_t$ depends on $\tau \setminus \kappa$. Putting this together, we average Eq.~\eqref{eq:normpresboundTtau} over all $2^N$ subsets $\tau$, getting:
\be
\label{eq:Peres}
\begin{split}
\Big|\Big|\big\langle T(A)\big\rangle\Big|\Big|^2 &\leq \Big\langle \prod_{k=1}^N\big|\big| A^{(k)}\big|\big|^2 \Big\rangle + %\\ &
2^{-N}\sum_{\tau,t} \Big\langle\big[A_{\sigma}^{(\kappa)},A_{\sigma'}^{(\kappa)}\big]\otimes R_t\Big\rangle (-1)^{|\tau\cap\kappa|}.
\end{split}
\ee
But the second line in Eq.~\eqref{eq:Peres} vanishes since
\be
\sum_\tau R_t(\tau\setminus\kappa) (-1)^{|\tau\cap\kappa|}=0,
\ee
as for every set $\tau\setminus\kappa$ there is one with and one without $\kappa$ in the sum, which mutually cancel due to the different signs. Hence, we recover the original inequality~\eqref{eq:VB}, completing the proof that no PPT state can give a violation.

We finish this section considering the extension of this result to a certain type of contractions:
\begin{corollary}[Peres' conjecture]
Let $T$ be a multi-linear contraction which is (i) a convex combination of norm-preserving maps or (ii) has a norm-preserving dilation (see Sec.~\ref{sec:Contractions}). Then Peres' conjecture holds as in Prop.\ref{prop:Peres}.
\end{corollary}
\proof{The fact that the validity of the inequality is stable w.r.t.\ both convex combinations and dilations follows easily from the triangle inequality of the norm.}

\subsection{Bipartite case}

Making use of a general result on moments whose proof and general statement we defer to section~\ref{sec:MomentInequalities}, we here prove the following:
\begin{proposition}[No Violation for $N=2$]
Bell inequalities arising from contractive maps are never violated for 2 parties.
\end{proposition}

In order to show this, we start by writing Bell inequalities~\eqref{eq:VB} for $N=2$:
\begin{equation}
\label{BellN2}
\sum_i \Big|\sum_{k,l} T_{i,kl} \ave{A_k \otimes B_l} \Big|^2 \leq \sum_{k,l} \ave{A_k^2 \otimes B_l^2}.
\end{equation}

We now diagonalize the $M\times M$ real matrix with elements $\Gamma_{kl}\equiv\ave{A_k \otimes B_l}$ as $O \Gamma U = \Lambda$, where $\Lambda$ is diagonal with nonnegative entries and $O$ and $U$ are orthogonal, namely,
\begin{equation}
%\begin{split}
\label{diagonalT}
 \sum_i \Big|\sum_{k,l} T_{i,kl} \ave{A_k \otimes B_l} \Big|^2 =%\\
%& \sum_i \Big|\sum_{k,l} \sum_{n,m,\alpha,\beta} O^{-1}_{n\alpha}O_{\alpha k} \ave{A_k \otimes B_l} U_{l\beta} U^{-1}_{\beta m} T_{i,nm} \Big|^2 =\\
%& \sum_i \Big|\sum_{\alpha,\beta} \delta_{\alpha \beta} \ave{\underbrace{\sum_k O_{\alpha k} A_k}_{\widetilde{A}_{\alpha}} \otimes \underbrace{\sum_l U_{l\beta} B_l}_{\widetilde{B}_{\beta}}  } \underbrace{\sum_{n,m} O_{\alpha n} T_{i,nm} U_{m \beta}}_{\widetilde{T}_{i,\alpha \beta}} \Big|^2=\\
 \sum_i \Big|\sum_{\alpha} \widetilde{T}_{i,\alpha \alpha} \ave{\widetilde{A}_{\alpha} \otimes \widetilde{B}_{\alpha}} \Big|^2,
%\end{split}
\end{equation}
where we have defined:
\be
\begin{split}
\widetilde{A}_{\alpha}&=\sum_k O_{\alpha k} A_k,\\
\widetilde{B}_{\beta}&=\sum_l U_{l\beta} B_l, \quad \textrm{and}\\
\widetilde{T}_{i,\alpha \beta}&=\sum_{n,m} O_{\alpha n} T_{i,nm} U_{m \beta},
\end{split}
\ee
and we have made use of the orthogonality of $O$ and $U$.

Choosing vectors $a^{\gamma}$ and $b^{\gamma}$ with components $(a^{\gamma})_i=\delta_{i\gamma}\langle\widetilde{A}_{\gamma} \otimes \widetilde{B}_{\gamma}\rangle$ and $(b^{\gamma})_i=\delta_{i\gamma}$, where $\delta_{i\gamma}$ is Kronecker's delta, we have:
\begin{equation}
\begin{split}
\label{diagToNorm}
&\norm{\sum_{\gamma}\widetilde{T}(a^{\gamma}, b^{\gamma})}^2=%\\
%&\sum_i \Big| \sum_{\gamma}\widetilde{T}_i(a^{\gamma}, b^{\gamma}) \Big|^2=\\
%&\sum_i \Big| \sum_{\gamma} \sum_{k,l} \widetilde{T}_{i,kl} \ave{\widetilde{A}_{\gamma} \otimes \widetilde{B}_{\gamma}} \delta_{k \gamma} \delta_{l \gamma} \Big|^2=\\
 \sum_i \Big|\sum_{\gamma} \widetilde{T}_{i,\gamma \gamma} \ave{\widetilde{A}_{\gamma} \otimes \widetilde{B}_{\gamma}} \Big|^2.
\end{split}
\end{equation}

Using then Eqs.~\eqref{diagonalT} and~\eqref{diagToNorm}, we can bound the l.h.s\ of the Bell inequality~\eqref{BellN2} as:
\begin{equation}
\begin{split}
&\sum_i \Big|\sum_{k,l} T_{i,kl} \ave{A_k \otimes B_l} \Big|^2 =
%\sum_i  \Big|\sum_{\gamma}  \widetilde{T}_{i,\gamma \gamma} \ave{\widetilde{A}_{\gamma} \otimes \widetilde{B}_{\gamma}} \Big|^2 =  \\
\norm{\sum_{\gamma}\widetilde{T}(a^{\gamma}, b^{\gamma})}^2 \leq \\
&\leq \left( \sum_{\gamma} \norm{\widetilde{T}(a^{\gamma},b^{\gamma})} \right)^2 \leq
\left( \sum_{\gamma} \norm{a^{\gamma}} \norm{b^{\gamma}} \right)^2 = %\\&
\left( \sum_{\gamma} \abs{\ave{\widetilde{A}_{\gamma} \otimes \widetilde{B}_{\gamma}}} \right)^2,
\end{split}
\end{equation}
where the two inequalities follow, respectively, from the triangle inequality for the norm and the contractivity of $\widetilde{T}$ (which easily follows from the contractivity of the original $T$), and in the last step we inserted the values of the norms of $a^{\gamma}$ and $b^{\gamma}$.% given in Eq.~\eqref{abnorms}.

We now get rid of the absolute value by noting that we can actually work with positive operators by making the decomposition:
$\tilde{A}_k = P_k - Q_k$,
where $P_k$ and $Q_k$ are the positive operators containing respectively the positive and negative eigenvalues of $\tilde{A}_k$. Thus, we can substitute the $\tilde{A}_k$ by $\tilde{A}'_k = P_k +Q_k$ and similarly for the $\tilde{B}_l$ so that
\begin{equation}
\left( \sum_{\gamma} \abs{\ave{\widetilde{A}_{\gamma} \otimes \widetilde{B}_{\gamma}}} \right)^2 \leq
\left( \sum_{\gamma} \ave{\widetilde{A}'_{\gamma} \otimes \widetilde{B}'_{\gamma}} \right)^2.
\end{equation}

Summing up, we have that if
\begin{equation}
\label{Ando22}
\left( \sum_{\gamma} \ave{\widetilde{A}'_{\gamma} \otimes \widetilde{B}'_{\gamma}} \right)^2 \leq
\sum_{k,l} \ave{\widetilde{A}'^2_{k} \otimes \widetilde{B}'^2_l}
\end{equation}
holds for any positive $\widetilde{A}'$ and $\widetilde{B}'$, we recover the original Bell inequality Eq.~\eqref{BellN2}:
\begin{equation}
\begin{split}
\sum_i \Big|\sum_{k,l} T_{i,kl} \ave{A_k \otimes B_l} \Big|^2 \leq
\left( \sum_{\gamma} \ave{\widetilde{A}'_{\gamma} \otimes \widetilde{B}'_{\gamma}} \right)^2 \leq %\\
\sum_{k,l} \ave{\widetilde{A}'^2_{k} \otimes \widetilde{B}'^2_l}=
\sum_{k,l} \ave{A_k^2 \otimes B_l^2},
\end{split}
\end{equation}
where the last equality follows from the definition of the primed and tilde operators together with the orthogonality of $O$ and $U$. But relation~\eqref{Ando22} is a specialized form for $N=2$ and $p=2$ of the general relation Eq.~\eqref{finalAndo} proven in section~\ref{sec:MomentInequalities}, which then completes the proof that there can be no violation of the Bell inequalities in the bipartite case.

\subsection{Orthogonal transformations}

Finally, we consider in this section the case of orthogonal transformations. The simplest construction for a norm-preserving map amounts to choosing $D=M^N$ and take $T$ orthogonal, but this turns out to only yield trivial inequalities:
\begin{proposition}[No violation for orthogonal maps]\label{prop:useless2} Let $D=M^N$ and $T_{i,s}$ be the entries of an orthogonal matrix. Then Eq.(\ref{eq:VB}) is always valid, independent of whether the expectation values come from a LHV model, quantum mechanics or any other theory.
\end{proposition}
\proof{The proof is straightforward. By exploiting orthogonality we can rewrite Eq.(\ref{eq:VB})
as \be \sum_s\Big\langle\prod_k A_{s_k}^{(k)}\Big\rangle^2\leq \sum_s\Big\langle\prod_k \Big(A_{s_k}^{(k)}\Big)^2\Big\rangle \ee which is always true by convexity of the square (\ie, the variance inequality).}

%%%%%%%%%%%%%%%%%%%%%%%
%
%
\section{Constructing the Maps}
\label{sec:MapConstruction}

Now that we have established the Bell inequality framework and studied some of its applications and limitations, we consider the construction of the maps that will ultimately give rise to the inequalities. We first tackle the question of constructing norm-preserving maps, and then we discuss possible ways to obtain contractions from this kind of maps.

\subsection{Constructing norm-preserving maps}
\label{sec:NormPreserving}

As we have already mentioned, the norm-preserving maps yield good candidates for constructing useful Bell inequalities. In order to construct non-trivial (and in the end all) norm-preserving maps $T$ it is advantageous to consider the real and positive operator $P:=T^\dagger T$ for which ${\rm rank}(P)\leq D$. Denoting the set of all normalized real product vectors in ${\mathbb{R}^{M}}^{\otimes N}$ by $V$, $T$ being a contraction means \be\langle\psi|P|\psi\rangle\leq 1,\quad\forall\psi\in V,\label{eq:P}\ee with equality for norm-preserving $T$. Conversely, given a real positive $P$ satisfying Eq.~\eqref{eq:P} (with equality) we can always construct a contraction (norm-preserving map) \be\label{eq:TfomrP}T=\sum_{i=1}^{D}|i\rangle\langle\varphi_i|\ee from any orthonormal basis $\{|i\rangle\}$ and the spectral decomposition $P=\sum_{i=1}^D|\varphi_i\rangle\langle\varphi_i|$. 

Equivalently, we may start with real symmetric matrices $Q$ satisfying\be\label{eq:Q} \langle\psi|Q|\psi\rangle\geq 0 ,\quad\forall\psi\in V,\ee and then set $P=\1-c Q$ with $0\leq c\leq1/||Q||$. In the language of entanglement theory $Q$ is an entanglement witness albeit only for real product states. It is now simple to characterize all norm-preserving $T$'s via the respective $Q$'s. To this end we denote by $\{G_\gamma\}$ the set of $M^2$ real $M\times M$ generalized Gell-Mann matrices which consists out of symmetric matrices (say for $\gamma\geq 0$) and anti-symmetric matrices (for $\gamma<0$) and $G_0:=\1$. This constitutes an orthogonal basis w.r.t.\ the Hilbert-Schmidt scalar product with elements satisfying ${\rm tr}[G_\gamma G^T_{\gamma'}]=2\delta_{\gamma \gamma'}$ for $\gamma, \gamma' \neq 0$. This leads us to:
\begin{proposition}[All norm-preserving maps]
\label{prop:Q}
Let $Q$ be a real symmetric matrix. Then the following are equivalent:
\begin{enumerate}
  \item $\langle\psi|Q|\psi\rangle = 0 ,\quad\forall\psi\in V,$
  \item $\sum_{\tau} Q^{T_\tau}=0$ where $^{T_\tau}$ denotes the partial transposition and the sum runs over all $2^N$ subsets $\tau\subseteq\{1,\ldots, N\}$.
  \item $Q=\sum_{g} c_g \bigotimes_{k=1}^N G_{g_k}$, where $c_g\in\mathbb{R}$ is zero whenever all components of $g=(g_1,\ldots,g_N)$ are non-negative (which we denote by $g\geq0$).
\end{enumerate}
\end{proposition}
\proof{$1\leftrightarrow 3$ follows from the fact that the real linear span of $|\psi\rangle\langle\psi|$, $\psi\in V$ is that of $G_g:=\bigotimes_{k=1}^N G_{g_k}$ with $g\geq 0$ together with the fact that the $G$'s form an orthogonal basis. $3\rightarrow 2$ easily follows from each summand appearing with both signs and $2\rightarrow 1$ is implied by
$$2^N\langle\psi|Q|\psi\rangle=\sum_\tau \tr{Q^{T_\tau}|\psi\rangle\langle\psi|^{T_\tau}} = \sum_\tau \tr{Q^{T_\tau}|\psi\rangle\langle\psi|}$$.}

Although Prop.~\ref{prop:Q} provides a simple way of constructing all norm-preserving maps $T$, it does a priori not provide good control over $D$ and we will generically end up with $D\approx M^N$. In the following we will outline a specific construction leading to considerably smaller $D$. For simplicity we begin with $N=2$ for which Prop.~\ref{prop:Q} boils down to $Q^{T_1}=-Q$. Let $\F:=\sum_{j,l=1}^M|j,l\rangle\langle l,j|$ be the flip operator.
\begin{lemma}The matrix
 \be \label{eq:newP} P=\1-\F+\F^{T_1}=\1+\sum_{\gamma<0}G_\gamma\otimes G_\gamma\ee is positive semi-definite,
satisfies Eq.(\ref{eq:P}) with equality and has ${\rm rank}(P)=1+(M^2-M)/2$. For $M=2,3$ this is the smallest possible one.
\label{lemma}
\end{lemma}
\proof{ Observe that $(\1-\F)/2$ is the projector onto the anti-symmetric subspace (of dimension $(M^2-M)/2$) and $\F^{T_1}/M$ the projector onto the standard maximally entangled state. This implies positivity and ${\rm rank}(P)=1+(M^2-M)/2$ %D$.
That $\langle\psi|P|\psi\rangle =1$ for all $\psi\in V$ follows from Prop.~\ref{prop:Q} and that ${\rm rank}(P)=4$ is smallest for $M=3$ results form the impossibility of ${\rm rank}(P)=3$ (\eg, due to Hurwitz' theorem on composition algebras~\cite{Shapiro}). }

 %Together with Eq.(\ref{eq:TfomrP})
 This leads us to
 \begin{proposition}[Norm-preserving maps with small $D$]
\label{prop:newexp}
 $T_i=G_{1-i},\quad i=1,\ldots,D=1+(M^2-M)/2$ defines a norm-preserving map.
 \end{proposition}
 \proof{Let $|ab\>$ stand for an arbitrary normalized, real product vector
(omitting the tensor product between the vectors). We have to show
that $\<ab|T^\dagger T|ab\>=1$. Using that by definition
$\<i|T|ab\>=\<a|G_{1-i}|b\>$ this indeed follows from the previous
Lemma via
\bea  \<ab|T^\dagger T|ab\> &=& \sum_{i=1}^D \<ab|T^\dagger|i\>\<i|T|ab\> \\
&=& \<aa|P|bb\> = \<ab|P|ab\>,\eea
where the last step is implied by the symmetry $(P\F)^{T_1}\F=P$.}

The dilated cross product map introduced in Eq.~\eqref{eq:DilatedCrossProductMap} provides a simple example of this construction. For $N>2$ we can clearly follow similar lines, but we may as well construct new norm-preserving maps by concatenation, as mentioned before. This is indeed the process leading to the explicit violation presented in Sec.~\ref{sec:ExplicitViolation}.

\subsection{Contractions from norm-preserving maps}
\label{sec:Contractions}

There are two basic ways of obtaining contractions from norm-preserving maps: (i) taking convex combinations and (ii) truncating the output space. For the Euclidean norm and  $D=1, N=2$ (or equivalently $N=1$) it is well known that, on the one hand, every contraction can be obtained as a convex combination of norm-preserving maps and, on the other hand, \emph{dilated} to a norm-preserving map with $M'=2M$. For $D\geq 2$ and $N\geq 2$ the situation is more complicated: it is clear that norm-preserving maps are no longer the only extreme points of the set of contractions, simply because there are too few of them (or none at all, \eg, for $M=D=3$). The set of contractions which can be dilated is, however, still richer and this is what will be studied in the following.

\begin{proposition}[Contractions with dilations]
\label{prop:dilations}
Let $T:\mathbb{R}^{M}\times\ldots\times\mathbb{R}^{M}\rightarrow \mathbb{R}^D$ be an $N$-linear contraction w.r.t.\ the Euclidean norm. Then the following are equivalent:
\begin{enumerate}
    \item $T$ has a norm-preserving dilation, \ie, there is a norm-preserving map with output space of dimension $D'\geq D$ whose restriction onto $\mathbb{R}^D$ is $T$.
\item There is a $Q$ of the form in Prop.\ref{prop:Q} such that $\1\geq T^\dagger T+Q$.
\item $\tr{\rho T^\dagger T}\leq 1$ for all real density operators $\rho=\sum_{g\geq 0} c_g \bigotimes_{k=1}^n G_{g_k}$.
\end{enumerate}
\end{proposition}
\proof{For $1\rightarrow 2$ let us denote the dilation by $\tilde{T}$.
Then $\tilde{T}^\dagger\tilde{T}\geq T^\dagger T$ and by Prop.~\ref{prop:Q} $\tilde{T}^\dagger\tilde{T}=\1-Q$. Conversely ($2\rightarrow 1$) there are vectors $|\varphi_i\rangle$ such that $\1-Q=T^\dagger T+\sum_{i=D+1}^{D'}|\varphi_i\rangle\langle\varphi_i|$ and we can set $\tilde{T}=\sum_{i=1}^{D'} |i\rangle\langle\varphi_i|$ which becomes $T$ when the sum is truncated at $D$. For $2\leftrightarrow 3$ observe that both can be cast into a semidefinite program: minimize $x$ s.t.\ $x\1\geq T^\dagger T+\sum_{g\not\geq 0} c_g G_g$ is the dual of maximize $\tr{\rho T^\dagger T}$ over the constrained set of $\rho$'s. By duality the two programs give the same value (as the primal program is bounded and strictly feasible) which proves equivalence of 2 and 3. }

Note that Prop.~\ref{prop:dilations} in particular implies that there is an efficient and certifiable algorithm (a semi-definite program) for deciding whether or not a given contraction has a norm-preserving dilation.

%%%%%%%%%%%%%%%%%%%%%%%%%%%%%
%
%
\section{Generalizations to Other Domains and Norms}
\label{sec:Other}

Up to now, we have set and studied the framework using the Euclidean norm and unrestricted measurement outcomes. In general, we may use $p$-norms defined as $||a||_p=\big(\sum_{j} |a_j|^p\big)^{1/p}$. We may also restrict the range $\cal R$ of the input vector components and ask $T$ to be a contraction only w.r.t.\ $A^{(k)}\in{\cal R}^M\subseteq\mathbb{R}^{M}$, which amounts to restricting the possible measurement outcomes.

We start by restating Prop.~\ref{prop:VB} for any $p$-norm:
\begin{proposition}[Bell inequalities for $p$-norms]\label{prop:VBp} Let $T$ be a multi-linear contraction on
$\cal{R}^{M}\times\ldots\times\cal{R}^{M}$ w.r.t.\ any $p$-norm with $p\geq 1$. Then  every LHV model has to satisfy
\be\label{eq:VBp}  \Big|\Big|\big\langle T(A)\big\rangle\Big|\Big|_p^p \leq \Big\langle \prod_{k=1}^N\big|\big| A^{(k)}\big|\big|_p^p\Big\rangle
\ee
for all observables with measurement outcomes in $\cal{R}$.
\end{proposition}
This follows, as before, by taking expectation value of the $p$-th power of Eq.~\eqref{eq:TA} and exploiting the convexity of the power function, namely, H\"older's inequality~\cite{bhatia}:
\be
\label{eq:Holders}
\tr{ \abs{A}^{p} \rho} \geq \tr{ \abs{A} \rho}^{p}.
\ee

In the following, we explore the range of applicability of these generalized inequalities by first showing how no violation is possible from the $p=1$ norm, and subsequently how we recover all standard Bell inequalities from the $p=\infty$ norm. We further discuss how probability Bell inequalities can be encompassed in the framework.

\subsection{No violation from the $p=1$ norm}

When considering the $p=1$ norm, we are once again faced with an impossibility result:
\begin{proposition}[No violation for $p=1$]\label{prop:useless1} Let  $0\in {\cal R}$ and $p=1$, then inequality (\ref{eq:VBp}) is always fulfilled---independent of whether the expectation values come from a LHV model, quantum mechanics or any other theory.
\end{proposition}
\proof{
In this case $T$ being a contraction implies that $\sum_{i=1}^D |T_{i,s}|\leq 1$ for all $s\in\{1,\ldots,M\}^N$. Hence we can bound the r.h.s.\ of Eq.(\ref{eq:VBp}) by \bea \sum_s \Big\langle\Big|\prod_{k=1}^N A_{s_k}^{(k)}\Big|\Big\rangle &\geq& \sum_s \Big|\Big\langle\prod_{k=1}^N A_{s_k}^{(k)}\Big\rangle\Big| \\
&\geq& \sum_i\sum_s \Big| T_{i,s} \Big\langle\prod_{k=1}^N A_{s_k}^{(k)}\Big\rangle\Big|,\eea which is in turn larger than the l.h.s.\ of Eq.(\ref{eq:VBp}).}

\subsection{Standard Bell inequalities from the $p=\infty$ norm}

For discussing the case $p=\infty$ we will, for the sake of simplicity, assume $\cal R$ to be countable and finite, which clearly reflects the experimental situation. With some care this can easily be relaxed. Denote by $A$ the value of the product $\prod_k \big|A_{s_k}^{(k)}\big|$ and by $P(A|s)$ the probability of measuring $A$ in a given setting $s$.  For the r.h.s.\ of Eq.(\ref{eq:VBp}) taken to the $1/p$-th power we get then
\bea &&\hspace*{-7pt} \lim_{p\rightarrow\infty}\left(\sum_s\Big\langle \prod_{k=1}^N \big|A_{s_k}^{(k)}\big|^p \Big\rangle\right)^{1/p}\hspace{-7pt} = \lim_{p\rightarrow\infty} \left(\sum_{A,s} P(A|s) |A|^p\right)^{1/p}\nonumber =\\&& 
= \max\Big\{|A|\;\Big|\exists s:P(A|s)>0\Big\}=:A_{\max} ,\label{eq:maxA}
\eea
where Eq.(\ref{eq:maxA}) is obtained by bounding $P(A|s)$ by its maximal and minimal non-zero value from above and below, and then taking the limit $p\rightarrow\infty$.  Eq.(\ref{eq:VBp}) thus becomes \be \max_i\Big|\big\langle T(A)_i \big\rangle\Big|\leq A_{\max}\label{eq:standard}.\ee  Since the maximum over $i$ is taken anyhow, we can as well drop this index and consider multi-linear functionals $T:\mathbb{R}^M\times\ldots\times\mathbb{R}^M\rightarrow\mathbb{R}$ in the first place. Then Eq.(\ref{eq:standard}) is nothing but a standard linear correlation Bell inequality, reasonably considered for ${\cal R}=\{\pm 1\}$.

\begin{proposition}
[Linear correlation inequalities] For $p=\infty$ the inequalities of the form in Eq.(\ref{eq:VBp}) are precisely those obtained in the framework of linear correlation Bell inequalities.
\end{proposition}
Note that they correspond to faces of the correlation polytope with dichotomic observables iff $T$ is an extreme point within the set of contractions.

\subsection{General Bell inequalities}

The restrictions in the range also allow dealing with general Bell inequalities, not necessarily involving correlation operators. In order to deal with this scenario, we restrict the input vectors in such a way that each component is associated to one of the possible measurement results. These components can only be equal to one or zero, depending on the outcome distribution predicted by a deterministic LHV model: $A^{(k)}\in{\cal R}^M$, where now ${\cal R}=\{ v \in \{0,1\}^{n} | \sum_k v_k =1 \}\subseteq \mathbb{R}_+^n$, where $n$ is the number of measurement outcomes. The range of possible input vectors is then rather limited: all the input vectors are such that whenever one of the components is one, the rest of components for the same measurement have to be zero. For instance, in the case of two measurements, $A_1$ and $A_2$, of two outcomes, $\pm 1$, each party has the four-dimensional vector, $(A_{1,+},A_{1,-},A_{2,+},A_{2,-})$. These components are either one or zero and such that $A_{1,+}+A_{1,-}=A_{2,+}+A_{2,-}=1$. This process specifies the range of input vectors ${\cal R}$ on which the contraction applies.

Now, in order to build the contraction, note that it is always possible to define the inequality in such a way that the maximum and minimum value for LHV models have the same absolute value. Now, we can adjust the norms and interpret the Bell inequality as a contraction over the restricted input vectors. This is nothing but a reformulation of the standard derivation of the LHV bound for LHV models.

The reformulation as a contraction however has some advantages. First, it allows generating inequalities for larger number of parties starting from the bipartite case by concatenating different contractions. Second, whenever the operator $\mathcal O$, defined in~\eqref{eq:odef}, consists only of commutators, the inequality satisfies the Peres' conjecture, as our proof straightforwardly applies to this situation. Note that in this case, and since the range of the input vectors is limited, one cannot conclude that Eq.~\eqref{eq:odef} automatically holds for norm-preserving maps.

Finally, it is worth mentioning that the Clauser-Horne-Shimony-Holt~\cite{CHSH} and, more in general, all the Mermin~\cite{Mermin} inequalities can be seen as contractions over input vectors whose components are between $-1$ and $+1$. Moreover, for these inequalities, the operator $\mathcal O$ only consists of commutators. This fact was indeed exploited in~\cite{WW00} to prove that all Mermin inequalities satisfy the Peres' conjecture.

%%%%%%%%%%%%%%%%%%
%
%
\section{Moment Inequalities}
\label{sec:MomentInequalities}

In this section, we abandon Bell inequalities for a while and prove a general result on moments. For this, we make use of Ando's theorem on joint concavity, which can be stated as follows~\cite{ando}:
Let $r_1, r_2, \ldots, r_N$ be positive numbers satisfying $r_1+\cdots+r_N\leq1$. Then the product $X^{(1)^{r_1}} \otimes X^{(2)^{r_2}} \otimes \cdots \otimes X^{(N)^{r_N}}$  is jointly concave on $N$-tuples of positive operators $X^{(1)},  \ldots, X^{(N)}$. In other words, given $X^{(j)}=\sum_\alpha \lambda_\alpha X^{(j)}_\alpha$ the same convex combination of positive operators $X^{(j)}_\alpha$ for all $j$, we have that:
\begin{equation}
\label{ando}
\sum_\alpha \lambda_\alpha \left(X_\alpha^{(1)^{r_1}} \otimes \cdots \otimes X^{(N)^{r_N}}_\alpha\right) \leq X^{(1)^{r_1}} \otimes \cdots \otimes X^{(N)^{r_N}},
\end{equation}
where inequality $X\leq Y$ among operators denotes that the operator $Y-X$ is positive semidefinite.

With this result in hand, we proceed to prove the following statement about correlations:
\begin{proposition}[Moment inequality]\label{prop:MomentInequality}
Let $A_k^{(j)}$ be positive operators and $M, N \in {\mathbb N}$ and $p\in {\mathbb R}$ positive numbers with $N \leq p$. Then the following holds:% for any state $\rho$:
\begin{equation}
\label{finalAndo}
%\abs{\sum_{k=1}^M \tr{\bigotimes_{j=1}^N A_k^{(j)} \rho}}^p \leq M^{(p-N)} \! \! \! \! \sum_{k_1,\ldots, k_N =1}^M \tr{\bigotimes_{j=1}^N \abs{A_{k_j}^{(j)}}^p \rho}.
\abs{\sum_{k=1}^M \ave{\bigotimes_{j=1}^N A_k^{(j)} }}^p \leq M^{(p-N)} \! \! \! \! \sum_{k_1,\ldots, k_N =1}^M \ave{\bigotimes_{j=1}^N \big( A_{k_j}^{(j)}\big)^p}.
\end{equation}
\end{proposition}

In order to show this, we start by extending the space as follows (we drop summation and product limits for clarity):
\begin{equation}
%\begin{split}
\label{extendedSpace}
\sum_k \ave{\bigotimes_{j} A_k^{(j)} }=% \sum_k \tr{\bigotimes_j A_k^{(j)} \rho} = \\
%&\tr{  \sum_k \bigg( A^{(1)}_{k} \otimes  A^{(2)}_{k}  \otimes \cdots \otimes  A^{(N)}_{k} \bigg) \rho }=\\
%&\tr{ \bigg(\sum_{k_1} A^{(1)}_{k_1} \otimes \ketbra{k_1} \bigg) \otimes \bigg( \sum_{k_2} A^{(2)}_{k_2}\otimes \ketbra{k_2} \bigg) \otimes \cdots \right.\\
%&\left. \cdots \otimes \bigg(\sum_{k_N} A^{(N)}_{k_N}\otimes \ketbra{k_N} \bigg) \; \bigg( \rho \otimes \sum_{s,t} \ket{s\ldots s}\bra{t\ldots t} \bigg)}\\
\tr{ \bigotimes_j \bigg( \sum_{k_j} A^{(j)}_{k_j} \otimes \ketbra{k_j} \bigg) \tilde{\rho} },
%\end{split}
\end{equation}
where we have defined $\tilde{\rho}=\rho\otimes\sum_{s,t} \ket{s\ldots s}\bra{t\ldots t}$ and $\rho$ is the original state over which the expectation values in~\eqref{finalAndo} are taken. The GHZ-like state $\sum_{s,t} \ket{s\ldots s}\bra{t\ldots t}$ we introduced in the extended space takes care of contracting the indices so that we recover the single index summation we had before. Note that the factors that appear are block matrices containing all $M$ $A_k^{(j)}$ operators for the given site $j$.

We now introduce unitary shift operators of the form:
\begin{equation}
S=\sum_k \1 \otimes \ket{k}\bra{k+1},
\end{equation}
where the addition inside the bra is understood to be modulo $M$. These operators act on each site as:
\begin{equation}
S \bigg( \sum_k A_k^{(j)} \otimes \ketbra{k} \bigg) S^\dagger = \sum_k A^{(j)}_{k+1} \otimes \ketbra{k}.
\end{equation}
That is, they shift (cyclically) by one position the blocks in the matrix. Applying the shift $\alpha$ times one simply gets:
\begin{equation}
S^\alpha \bigg( \sum_k A_k^{(j)} \otimes \ketbra{k} \bigg) S^{\dagger\alpha} = \sum_k A^{(j)}_{k+\alpha} \otimes \ketbra{k}.
\end{equation}
Going back to expression~\eqref{extendedSpace}, we can shift by the same amount in all sites without changing the overall value, so that we can work with the average
\begin{equation}
%\begin{split}
\sum_k \ave{\bigotimes_{j} A_k^{(j)} }=
\frac{1}{M} \sum_\alpha \tr{ \bigotimes_j S^\alpha \bigg(\sum_{k_j} C^{(j)^{1/p}}_{k_j} \otimes \ketbra{k_j} \bigg) S^{\dagger \alpha} \; \tilde{\rho}  },
%&\sum_k \tr{\bigotimes_j A_k^{(j)} \rho} = \\
%&\frac{1}{M} \sum_\alpha \tr{ S^\alpha \bigg(\sum_{k_1} A^{(1)}_{k_1} \otimes \ketbra{k_1} \bigg) S^{\dagger \alpha} \otimes   \cdots \right.\\
%&\left. S^\alpha \bigg( \sum_{k_2} A^{(2)}_{k_2}\otimes \ketbra{k_2} \bigg) S^{\dagger \alpha} \otimes \cdots \right. \\
%&\left. \cdots \otimes S^\alpha \bigg(\sum_{k_N} A^{(N)}_{k_N}\otimes \ketbra{k_N} \bigg) S^{\dagger \alpha} \;\; \tilde{\rho} }.%\right. \\
%&\left. \bigg( \rho \otimes \sum_{s,t} \ket{s\ldots s}\bra{t\ldots t} \bigg)  }.
%\end{split}
\end{equation}
where we have defined $C_k^{(j)}=\big(A_k^{(j)}\big)^p$.

We can now bring the $1/p$ powers outside of the block matrices and, using the unitarity of the $S$, we get:
\begin{equation}
\begin{split}
\label{preAndo}
&\sum_k \ave{\bigotimes_{j} A_k^{(j)} }= %\\&
%&\sum_k \tr{\bigotimes_j A_k^{(j)} \rho} =\\
%& \frac{1}{M} \sum_\alpha \tr{ \bigotimes_j \bigg[ S^\alpha \bigg(\sum_{k_j} C^{(j)}_{k_j} \otimes \ketbra{k_j} \bigg) S^{\dagger \alpha} \bigg] ^{1/p}\; \tilde{\rho}  }=\\
\frac{1}{M} \sum_\alpha \tr{ \bigotimes_j  \bigg(\sum_{k_j} C^{(j)}_{k_j+\alpha} \otimes \ketbra{k_j} \bigg)^{1/p} \; \tilde{\rho}  }.
\end{split}
\end{equation}

We are now in a position to use Ando's result by identifying:
\begin{equation}
X^{(j)}_\alpha=\sum_{k_j} C^{(j)}_{k_j+\alpha} \otimes \ketbra{k_j},
\end{equation}
and hence
\begin{equation}
%\begin{split}
X^{(j)}=\frac{1}{M}\sum_\alpha \sum_{k_j} C^{(j)}_{k_j+\alpha} \otimes \ketbra{k_j}
%&=\frac{1}{M} \sum_\alpha C^{(j)}_\alpha \otimes \1
= \bar{C}^{(j)} \otimes \1,
%\end{split}
\end{equation}
where we defined the averaged operators
\begin{equation}
\bar{C}^{(j)}=\frac{1}{M} \sum_\alpha C^{(j)}_\alpha.
\end{equation}
Ando's theorem Eq.~\eqref{ando} then tells us that:
\begin{equation}
\frac{1}{M} \sum_\alpha \bigotimes_j  \bigg(\sum_{k_j} C^{(j)}_{k_j+\alpha} \otimes \ketbra{k_j} \bigg)^{1/p} \leq
\bigotimes_j \big( \bar{C}^{(j)}\otimes \1 \big)^{1/p}
\end{equation}
whenever $N/p \leq 1$. Plugging this in expression~\eqref{preAndo} and making use of H\"older's inequality~\eqref{eq:Holders}, we obtain:
\begin{equation}
\begin{split}
&\sum_k \ave{\bigotimes_{j} A_k^{(j)} }\leq
%\tr{ \bigotimes_j \big( \bar{C}^{(j)}\otimes \mathbf{1} \big)^{1/p}  \tilde{\rho}}
%\bigg(\rho \otimes \sum_{s,t} \ket{s\ldots s}\bra{t\ldots t} \bigg)}=\\
M \ave{ \bigotimes_j \big( \bar{C}^{(j)} \big)^{1/p}  } \leq %\\&
%M \Bigg( \ave{ \bigotimes_j \bar{C}^{(j)}  } \Bigg)^{1/p}=\\
M \Bigg( \ave{ \bigotimes_j \frac{1}{M} \sum_{\alpha_j} C^{(j)}_{\alpha_j}  } \Bigg)^{1/p} =\\&
=\frac{M}{M^{N/p}} \Bigg( \sum_{\alpha_1,\ldots,\alpha_N} \ave{ \bigotimes_j  \bigg(A^{(j)}_{\alpha_j}\bigg)^p } \Bigg)^{1/p},
\end{split}
\end{equation}
which completes the proof.

%%%%%%%%%%%%%%%%%%
%
%
\section{Discussion}
\label{sec:Discussion}

The present work arose out of the aim to better understand and generalize the non-linear inequalities presented in \cite{CFRD,SV}. The framework into which our work embeds these inequalities appears to be general and intuitive. In fact, it allows to embrace all standard Bell inequalities (as sketched in Sec.~\ref{sec:Other}) albeit there seems to be a trade-off between the appeal of the method and its generality.
Most interesting results seem to arise in the natural formulation involving the Euclidean norm. A relevant example is the proof of Peres' conjecture, which we have only presented for this basic case, and for norm-preserving maps. One can easily extend the proof to the case of $p$-norms with even $p$, but the extension to arbitrary $p$ and contractive maps seems trickier. Nevertheless, the idea of the proof could be more general than the present results about Bell inequalities not violated by PPT quantum states.

At the moment, the power of the framework is still to be explored. On the one hand, previously existing non-linear moment Bell inequalities are nicely encompassed and novel inequalities can be produced. On the other hand,
their potential usefulness arising from their validity for arbitrary observables is shadowed by our result on the impossibility of violations in the bipartite case. For three parties we saw that unbounded violations are possible, however, so far the result is more a proof of existence rather than a concrete recipe. We envisage possible future applications of the present results concerning the Peres' conjecture: whenever one can prove that a Bell inequality fits the conditions of proposition 2 (or corollary 1), it must satisfy the conjecture. Finally, we want to stress that the general moment inequality derived in Section VII might also find potential applications outside the realm of Bell inequalities.\vspace*{5pt}

{\bf Acknowledgements} The authors acknowledge financial support by the European project COQUIT under FET-Open grant number 233747,
QUANTOP, the Danish Research Council (FNU), the Spanish projects QTIT (FIS2007-60182) and QOIT
(Consolider Ingenio 2010), EU Integrated Projects QAP, ERC Grants PERCENT, Caixa Manresa,
Generalitat de Catalunya and Spanish grants I-MATH, MTM2008-01366 and CCG08-UCM/ESP-4394.

\end{document}